# A Dynamical Systems Approach to Predicting Patient Outcome after Cardiac Arrest

Richard J Povinelli[1], Mathew Dupont[2]

[1]Marquette University, Milwaukee, Wisconsin, U.S.
[2]Milwaukee, Wisconsin, U.S.

## Abstract

*Aim*: Approximately six million people suffer cardiac arrests worldwide per year with very low survival rates (<1%). Thus, the aim of this study is to estimate the probability of a poor outcome after cardiac arrest. Accurate outcome predictions avoid removing care too soon for patients with potentially good outcomes or continuing care for patients with likely poor outcomes.

*Method*: The method is based on dynamical systems embedding theorems that show that a reconstructed phase space (RPS) topologically equivalent to an underlying system can be constructed from measured signals. Here the underlying system is the human brain after a cardiac arrest, and the signals are the EEG channels. We model the RPS with a Gaussian mixture model (GMM) and ensemble the output of the RPS-GMM with clinical data via XGBoost.

*Results*: As team Blue and Gold in the Predicting Neurological Recovery from Coma After Cardiac Arrest: The George B. Moody PhysioNet Challenge 2023, our RPS-GMM-XGBoost method obtained a test set competition score of 0.426 and rank of 24/36.

## 1. Introduction

Of the six million cardiac arrests worldwide each year, less than 1% survive [1]. Thus, the objective of the Predicting Neurological Recovery from Coma After Cardiac Arrest: The George B. Moody PhysioNet Challenge 2023 is to predict patient outcomes after cardiac arrest [2]–[4], with a particular emphasis on finding methods that have low false positive rates; that is, we wish to avoid predicting a poor outcome when the patient will actually have a good outcome. Specifically, the challenge score is the true positive rate (TPR) when the false positive rate (FPR) was 0.05.

The contributions of this paper are the application of a reconstructed phase space (RPS) Gaussian mixture model (GMM) XGBoost approach to analyze EEG signals for cardiac arrest patient outcome prediction.

## 2. Literature Review

There has been substantial research in predicting patient outcomes post-cardiac arrest. Examples include the work of Scarpino et al. who predicted outcomes after 12 and 72 hours [5], Amorim et al. who used EEGs to predict poor functional outcome [6], and Tjepkema-Cloostermans et al. who used deep learning models to determine patient outcomes [7]. RPS-GMM based methods have been previously used to classify ECG arrhythmias [8].

## 3. Method

We approach this problem from a dynamical systems perspective. Thus, the hypothesis posed here is whether a dynamical system model of the brain can accurately predict cardiac arrest outcomes. Specifically, we use a RPS-GMM approach [9] that is ensembled with clinical data via XGBoost.

### 3.1. Preprocessing

Because of the large size of the training dataset (1.5 TB), it was not feasible to train the RPS-GMM-XGBoost on all the data. Thus, while the challenge dataset contained EEG, ECG, and other signals (SpO2, EMG, etc.), the RPS-GMM-XGBoost method used only the EEG signals. Similarly, the training data set had up to 72 hours (and sometimes more) of EEG signals, but we used the data from just hours 12, 24, 48, and 72. If a particular hour was missing, we used the next closest hour less than the desired hour, but greater than the next lowest hour. The RPS-GMM-XGBoost method is robust against missing data, so training with missing data is handled appropriately as is classifying a patient with missing data.

The full hour of the EEG signals is band-pass filtered with cutoff frequencies of 0.1 and 50Hz using a 10$^{th}$ order Butterworth filter. Then, all EEG channels are re-referenced to the average of the channels. Next, as the signals have varying sampling frequencies, all signals are down sampled to 100Hz using Fourier resampling.

The data set is further reduced to the last five minutes of each hour. Finally, a subset of 10 channels are used as illustrated below in Figure 1.

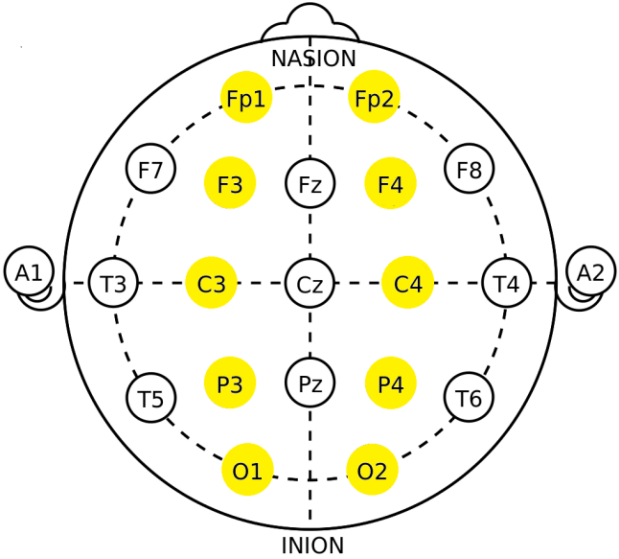

Figure 1. Electrode locations used by the RPS-GMM-XGBoost algorithm are highlighted in yellow.

### 3.2. RPS-GMM Method

RPS theory is based on the work of Takens [10], who showed that a space topologically equivalent to the phase space of a system can be constructed from one measured signal. Requirements for this theory are that the RPS be greater than twice the dimension of the original system and that the measured signal be twice continuously differentiable. Sauer et al. [11] extended Takens work by showing that the RPS need only be twice the dimension of the attractor in the original system's phase space and that the measured signal be only once continuously differentiable. The systems of interest such as the brain have very large dimensions, thus according to the Takens theorem the RPS would need to also be twice a very large dimension. In practice, lower dimensional RPSs will have self-intersections of the phase space trajectories but can still be effectively modelled using the techniques described next.

The RPS-GMM method constructs a matrix from a signal. Let $x$ be a signal, $x_n$ be the $n^{th}$ sample from $x$, and $N$ be the number of samples. Furthermore, let $d$ be the dimension of the RPS and $\tau$ be the lag between samples. The RPS is an $N-(d-1)\tau$ by $d$ matrix defined as

$$\mathbf{X} = \begin{bmatrix} x_1 & x_{1+\tau} & \cdots & x_{1+(d-1)\tau} \\ x_2 & x_{2+\tau} & \cdots & x_{2+(d-1)\tau} \\ \vdots & \vdots & \ddots & \vdots \\ x_{N-(d-1)\tau} & x_{N-(d-2)\tau} & \cdots & x_N \end{bmatrix}. \quad (1)$$

A point in the RPS is defined as

$$\mathbf{x}_n = \begin{bmatrix} x_n & x_{n+\tau} & \cdots & x_{n+(d-1)\tau} \end{bmatrix}. \quad (2)$$

In this work, we use a dimension of four and a time lag of 12.

The distribution of the RPS is modelled with a GMM. Let $\boldsymbol{\mu}$ be the center of one Gaussian mixture, $\Sigma$ be the covariance matrix describing the covariance between dimensions, and $w_m$ be the weight associated with one mixture. Let $N$ be a Gaussian distribution, $m$ be the index of a single Gaussian, and $M$ be the number of mixtures. A GMM is defined as

$$w_m N(\boldsymbol{\mu}_m, \boldsymbol{\Sigma}_m), m = 1 \ldots M. \quad (3)$$

There are three types of GMMs – spherical, diagonal, and full. The type of GMM is determined by the structure of the covariance matrix $\Sigma$. A full covariance GMM is a symmetric matrix where the off-diagonal values represent the covariance between different dimensions. We used 16 full covariance mixtures in this work.

To capture spatial and temporal features of the dataset, individual RPS-GMM models are trained. A GMM is estimated for each selected hour, channel, and outcome, i.e., four hours by 10 channels by two classes yields 80 GMMs. For each training sample in each group, an RPS is generated. All RPSs belonging to a particular group are then concatenated together into a combined data matrix. A 16 full covariance mixture GMM model is estimated using expectation maximization over the combined data matrix.

To classify a test signal, an RPS is formed. The probability of a point in the RPS being generated by a GMM is

$$\hat{p}(\mathbf{x}_n) = \sum_{m=1}^{M} w_m N(\mathbf{x}_n; \boldsymbol{\mu}_m, \boldsymbol{\Sigma}_m). \quad (4)$$

This is repeated for each of the 80 GMMs, i.e., each hour, channel, and outcome.

The log likelihood of the test RPS is calculated by adding the log likelihood for each hour and channel for a particular class. Let $c$ correspond to a channel, $C$ be the number of channels, $h$ correspond to an hour, $H$ be the number of hours, $n$ be a point in the RPS, and $N$ be the total number of points in the RPS. Then the likelihood of outcome $o$ is

$$\hat{p}_o = \sum_{c=1}^{C} \sum_{h=1}^{H} \log \hat{p}(\mathbf{X}_{c,h})$$
$$= \sum_{c=1}^{C} \sum_{h=1}^{H} \sum_{n=1}^{N} \log \hat{p}(\mathbf{x}_{c,h,n}) \quad . \quad (5)$$

The outcome is determined by maximum likelihood.

$$\hat{o} = \underset{o \in \{good, poor\}}{\arg\max} \{\hat{p}_o\} . \quad (6)$$

In summary, the RPS-GMM method forms a concatenated reconstructed phase space from the signals for each outcome. This is done for each hour, each channel, and outcome yielding 80 RPSs. A 16-mixture full covariance GMM is estimated on each RPS yielding 80 GMMs. To classify a test signal, RPSs for each hour and channel are formed. The probability of each point in all RPSs is estimated. For each outcome, the sum of the log probabilities is calculated. The outcome with the greatest sum is selected.

### 3.3. Clinical Data and Ensemble

An XGBoost model, with default hyperparameters, is used to incorporate clinical data and ensemble RPS-GMM models. XGBoost is used because of its training speed, strong performance across a broad set of problems, and tolerance of nan values for inputs. The XGBoost model is trained directly on the outcome and probability estimates from the trained RPS-GMM model concatenated with the clinical data including age, time to return of spontaneous circulation, sex, shockable rhythm, and whether the cardiac arrest occurred in or out of the hospital. Categorical clinical data is one-hot encoded.

### 3.4. Implementation

The RPS-GMM-XGBoost algorithm is implemented in Python (3.8.6). The GMM package is PyCave (3.2.1), which builds on PyTorch and enables GPU based training. The code to build the reconstructed phase space is based on the NoLiTSA package. Signal processing code is from SciPy (1.10.1).

### 4. Results

The RPS-GMM-XGBoost method is evaluated at hours 12, 24, 48, and 72 after the cardiac arrest. The scoring metric is based on the correct prediction of a good outcome while minimizing incorrect predictions of poor outcomes. This is the true positive rate when the false positive rate is 0.05 with adjustments to take into account hospital variability.

Table 1 presents the results of our RPS-GMM-XGBoost method.

| Hour | Training | Validation | Test | Ranking |
|---|---|---|---|---|
| 12 | 0.513 | 0.164 | 0.302 | 11/36 |
| 24 | 0.589 | 0.239 | 0.465 | 9/36 |
| 48 | 0.885 | 0.224 | 0.351 | 25/36 |
| 72 | 0.982 | 0.463 | 0.426 | 24/36 |

Table 1. True positive rate at a false positive rate of 0.05 for hours 12, 24, 48, and 72 for training, validation, and test data sets. Also included is the ranking for each task.

Our official challenge score and rank are 0.426 and 24/36. The training set is publicly available. We submitted two entries for validation. The validation score presented is for our best scoring method. The test set results were evaluated once. All results were provided by the competition organizers; therefore, the training results are in-sample.

The code to train and test our algorithm was submitted to the competition where it ran on a cloud based virtual machine. The code was allowed to run for 24 hours on a virtual machine with a GPU and 72 hours on a virtual machine with just a CPU. While the run time is unknown for the virtual machine, training on the 607 patients in the training set took 81 minutes, of which 46 minutes were loading and preprocessing the data, on a Windows 10 computer with an eight core 3.59GHz AMD Ryzen 3700X CPU, an NVIDIA GeForce RTX 3060 Ti GPU, and 64 GB of RAM. Some multiprocessing and GPU acceleration was unable to be incorporated into the official submission, so virtual machine training took significantly longer.

### 5. Discussion

The RPS-GMM-XGBoost algorithm has results as high as the 75[th] percentile rank for hour 24 and as low as the 30[th] percentile rank for hour 48. It is interesting to note that on the test set our algorithm would have performed better if we ignored all data after the 24[th] hour. Another interesting point to note is that our in-sample results for hour 72 are 0.982 or nearly perfect. Comparing this against both the validation and test sets strongly suggests that we are overfitting our model.

Given the size of the data set and computing resources we had available for model development, we used only a fraction of the data that is available. Model training used only 4 hours out of 72, 5 minutes out of 60, and 10 channels out of 19. We down sampled to 100Hz. Assuming that the average sampling rate was originally 300Hz, we used approximately 0.08% of the available training data. Even still more than half the training time is spent loading data. While it is likely that using more training data will improve results, we need to investigate why our training results are so much better than our test results and why our 24 hour results are better than our 48 and 72 hour results.

An alternative technique expanded on the features used in the XGBoost ensemble; it included as parameters the outcome probabilities of each of the 80 RPS-GMM models and added average band power signals for each of alpha, beta, delta, and theta waves for each channel and hour with frequencies 8.0-12.0Hz, 12.0-30.0Hz, 0.5-4.0Hz, 4.0-8.0Hz, respectively. Preliminary evaluation of the enhanced technique on the publicly available training set suggested improved performance over the scoring model with a challenge score of 0.571 on an 80/20 train/test split. This model was submitted for official scoring, but did not score.

An alternative preprocessing technique was evaluated. The least variant five minutes are selected from the chosen hour, by omitting any data points belonging to any half-second window with less than a threshold of 5 µV peak-to-peak variance, then selecting from the remaining data the contiguous 5-minute window with the least peak-to-peak variance. Experimentally, choosing the last five minutes performed better when used for the RPS-GMM-XGBoost model.

Three deep learning models were considered. The first model is a 1-dimensional convolutional neural network (CNN), convolving a kernel of length five with a stride of two across all channels. The second model consists of 25 Conv2DLSTM neurons in sequence. A final model first ran a 2-dimensional CNN model across all channels for each time point, for each hour; the weights for the CNN model are shared across each hour, and outputs from each hour's CNN model are then passed to an LSTM layer. The Conv2dLSTM and stacked CNN-LSTM models failed to converge. The 1D-CNN model converged but is computationally expensive and performed substantially worse than the baseline model.

## 6. Conclusion

The RPS-GMM-XGBoost method is able to predict patient outcomes at 72 hours with a TPR of 0.426 and at an FPR of 0.05. While this method is not the best method in the PhysioNet 2023 Challenge, it shows that dynamical system approaches should be considered as a component of a cardiac arrest outcome prediction system. Because of the large data set size and training time limits, we were unable to use the complete data set. Future work is to include more of the data set in the analysis and incorporate the RPS-GMM-XGBoost approach as part of an ensemble.

## References

[1] R. Mehra, "Global Public Health Problem of Sudden Cardiac Death," *Journal of Electrocardiology*, vol. 40, no. 6, Supplement 1, pp. S118–S122, 2007, doi: 10.1016/j.jelectrocard.2007.06.023.

[2] A. L. Goldberger *et al.*, "PhysioBank, PhysioToolkit, and PhysioNet," *Circulation*, vol. 101, no. 23, pp. e215–e220, Jun. 2000, doi: 10.1161/01.CIR.101.23.e215.

[3] E. Amorim *et al.*, "The International Cardiac Arrest Research (I-CARE) Consortium Electroencephalography Database," *Critical Care Medicine*, (in press), doi: 10.1097/CCM.0000000000006074.

[4] M. A. Reyna *et al.*, "Predicting Neurological Recovery from Coma after Cardiac Arrest: The George B. Moody PhysioNet Challenge 2023," in *Computing in Cardiology*, 2023, pp. 1–4.

[5] M. Scarpino *et al.*, "Neurophysiology for Predicting Good and Poor Neurological Outcome at 12 and 72 h after Cardiac Arrest: The ProNeCA Multicentre Prospective Study," *Resuscitation*, vol. 147, pp. 95–103, Feb. 2020, doi: 10.1016/j.resuscitation.2019.11.014.

[6] E. Amorim *et al.*, "Continuous EEG Monitoring Enhances Multimodal Outcome Prediction in Hypoxic-ischemic Brain Injury," *Resuscitation*, vol. 109, pp. 121–126, Dec. 2016, doi: 10.1016/j.resuscitation.2016.08.012.

[7] M. C. Tjepkema-Cloostermans *et al.*, "Outcome Prediction in Postanoxic Coma with Deep Learning," *Critical Care Medicine*, vol. 47, no. 10, 2019, doi: 10.1097/ccm.0000000000003854.

[8] F. M. Roberts, R. J. Povinelli, and K. M. Ropella, "Identification of ECG Arrhythmias using Phase Space Reconstruction," in *Principles and Practice of Knowledge Discovery in Databases (PKDD'01)*, L. de Raedt and A. Siebes, Eds., Freiburg, Germany: Springer Verlag, 2001, pp. 411–423. doi: 10.1007/3-540-44794-6.

[9] R. J. Povinelli, M. T. Johnson, A. C. Lindgren, F. M. Roberts, and J. Ye, "Statistical Models of Reconstructed Phase Spaces for Signal Classification," *IEEE Transactions on Signal Processing*, vol. 54, no., pp. 2178–2186, 2006, doi: 10.1109/TSP.2006.873479.

[10] F. Takens, "Detecting Strange Attractors in Turbulence," in *Dynamical Systems and Turbulence*, A. Dold and B. Eckman, Eds., Warwick: Springer-Verlag, 1981, pp. 366–381. doi: 10.1007/BFb0091924.

[11] T. Sauer, J. A. Yorke, and M. Casdagli, "Embedology," *Journal of Statistical Physics*, vol. 65, no. 3, pp. 579–616, 1991, doi: 10.1007/BF01053745.

Address and email for correspondence:

Richard J. Povinelli
1515 W. Wisconsin Ave.
Milwaukee, WI 53233, U.S.
richard.povinelli@marquette.edu